\documentstyle[11pt]{article}
\def\bea{\begin{eqnarray}}
\def\eea{\end{eqnarray}}

\def\beq{\begin{equation}}
\def\eeq{\end{equation}}
\def\ba{\beq\new\begin{array}{c}}
\def\ea{\end{array}\eeq}
\def\be{\ba}
\def\ee{\ea}

\makeatletter
\newdimen\normalarrayskip 
\newdimen\minarrayskip 
\normalarrayskip\baselineskip
\minarrayskip\jot
\newif\ifold \oldtrue \def\new{\oldfalse}
\def\arraymode{\ifold\relax\else\displaystyle\fi} 
\def\eqnumphantom{\phantom{(\theequation)}} 
\def\@arrayskip{\ifold\baselineskip\z@\lineskip\z@
\else
\baselineskip\minarrayskip\lineskip2\minarrayskip\fi}
\def\@arrayclassz{\ifcase \@lastchclass \@acolampacol \or
\@ampacol \or \or \or \@addamp \or
\@acolampacol \or \@firstampfalse \@acol \fi
\edef\@preamble{\@preamble
\ifcase \@chnum
\hfil$\relax\arraymode\@sharp$\hfil
\or $\relax\arraymode\@sharp$\hfil
\or \hfil$\relax\arraymode\@sharp$\fi}}
\def\@array[#1]#2{\setbox\@arstrutbox=\hbox{\vrule
height\arraystretch \ht\strutbox
depth\arraystretch \dp\strutbox
width\z@}\@mkpream{#2}\edef\@preamble{\halign
\noexpand\@halignto
\bgroup \tabskip\z@ \@arstrut \@preamble \tabskip\z@ \cr}%
\let\@startpbox\@@startpbox \let\@endpbox\@@endpbox
\if #1t\vtop \else \if#1b\vbox \else \vcenter \fi\fi
\bgroup \let\par\relax
\let\@sharp##\let\protect\relax
\@arrayskip\@preamble}
\def\eqnarray{\stepcounter{equation}%
\let\@currentlabel=\theequation
\global\@eqnswtrue
\global\@eqcnt\z@
\tabskip\@centering
\let\\=\@eqncr
$$%
\halign to \displaywidth\bgroup
\eqnumphantom\@eqnsel\hskip\@centering
$\displaystyle \tabskip\z@ {##}$%
\global\@eqcnt\@ne \hskip 2\arraycolsep
$\displaystyle\arraymode{##}$\hfil
\global\@eqcnt\tw@ \hskip 2\arraycolsep
$\displaystyle\tabskip\z@{##}$\hfil
\tabskip\@centering
&{##}\tabskip\z@\cr}
\begingroup\ifx\undefined\newsymbol \else\def\input#1 {\endgroup}\fi

\begin{document}

\setcounter{footnote}{1}
\def\thefootnote{\fnsymbol{footnote}}
\begin{center}
\hfill ITEP/TH-35/00\\
\hfill hep-th/0007072\\
\vspace{0.3in}
{\Large\bf On Relation Between Moyal and Kontsevich Quantum Products.
Direct Evaluation up to the ${\hbar}^{3}$-Order
}
\end{center}
\centerline{{\large A.Zotov}\footnote{
ITEP, Moscow, Russia and MIPT, Dolgoprudny, Russia; e-mail: zotov@gate.itep.ru}}

\bigskip

\abstract{\footnotesize
 In his celebrated paper Kontsevich has proved a theorem which
manifestly gives a quantum product (deformation quantization formula)
and states that changing coordinates
leads to gauge equivalent star products. To illuminate his procedure, we make an
arbitrary change of
coordinates in the Moyal product and obtain the deformation quantization formula
up to the third order. In this way, the Poisson bi-vector is shown to
depend on $\hbar$ and not to satisfy the Jacobi identity. It is also shown that
the values of coefficients in the formula obtained follow from associativity of
the star product.}

\begin{center}
\rule{5cm}{1pt}
\end{center}

\bigskip
\setcounter{footnote}{0}
\section{Introduction}

\paragraph{1.Star product.}

Let us introduce, following [1] a star product that is an associative
$R[[\hbar]]$ -- bilinear product on algebra $A[[\hbar]]$ of smooth functions
on a finite-dimensional $C^\infty$-manifold:

\be
f\star{g}=fg+\hbar{B_1(f,g)}+\hbar^2{B_2(f,g)}+\ldots\subset{A[[\hbar]]}
\ee
where $\hbar$ is a formal variable and $B_i(f,g)$ -- bidifferential operators.

Associativity for the $n^{th}$ order means:
$(f\star{g})\star{h}=f\star{(g\star{h})}+O(\hbar^{n+1}).$
\vskip 8mm
\paragraph{2.Gauge group.}

There is a natural gauge group which acts on star products:

\be
\star\rightarrow\star', f'(\hbar)=Df(\hbar),f'\star{'g'}=
D(D^{-1}f'\star{D^{-1}g'})
\ee
where $D=1+\sum_{i\geq{1}}{\hbar^{i}D_i}$ and $D_i$'s are arbitrary
differential operators

$$
f\rightarrow{f+\hbar{D_1}f+\hbar^{2}D_2f+O(\hbar^{3})}
$$
Associativity of the new star product is obvious since
$f'\star{'g'\star{'h'}}=D(f\star{g\star{h}})$.

It follows from above

\be
{B_1}'(f,g)=B_1(f,g)+D_1(fg)-fD_1(g)-gD_1(f)
\ee
$B_1(f,g)$ can be chosen to be a skew-symmetric bi-vector field (see [1]).
Then, we put
$B_1(f,g)=\alpha^{ab}\partial_a{f}\partial_b{g}$, $\alpha^{ab}=-\alpha^{ba}$.
The Poisson bi-vector may depend on $\hbar$:

\be
\alpha^{ab}(\hbar)=\sum\limits_{i\geq{0}}{\hbar}^{i}\alpha_i^{ab}
\ee
The second order term $O(\hbar^2)$ in the associativity equation
$f\star(g\star{h})=(f\star{g})\star{h}$ implies that $\alpha$
gives a Poisson structure on $X$,

$$
\forall f,g,h\ \ \ \  \{f\{g,h\}\}+\{g,\{h,f\}\}+\{h,\{f,g\}\}=0,
$$
where $\{f,g\}:=\frac{f\star{g}-g\star{f}}{\hbar}|_{\hbar=0}$

\vskip 8mm
\paragraph{3.Moyal product.}

An example of the star product is the Moyal product:

\be
f\star{g}=fg+\hbar\vartheta^{ij}\partial_i{f}\partial_j{g}+\\
\\+\frac{\hbar^{2}}{2!}\vartheta^{ij}\vartheta^{kl}\partial_i\partial_k{f}
\partial_j\partial_l{g}+
\frac{\hbar^{3}}{3!}\vartheta^{ij}\vartheta^{kl}\vartheta^{mn}
\partial_i\partial_k\partial_m{f}\partial_j\partial_l\partial_n{g}+\ldots=\\
=\sum\limits_{n=0}\limits^{\infty}\frac{\hbar^n}{n!}\sum\limits_{i_1,...,i_n;
j_1,...,j_n}\prod\limits^{n}\limits_{k=1}\vartheta^{i_k{j_k}}(\prod\limits_{k=1}
\limits^{n}\partial_{i_k})(f)
\times(\prod\limits_{k=1}\limits^{n}\partial_{j_k})(g)=\\
\left.=e^{\hbar\vartheta^{ij}
{\partial_i}^{(1)}{\partial_j}^{(2)}}f(x_{(1)})g(x_{(2)})
\right|_{x_{(1)}=x_{(2)}=x}
\ee
where $\vartheta^{ij}$ is constant and skew-symmetric.
\vskip 8mm
\paragraph{4. Kontsevich formula.}

In paper [1] the following theorem was stated

\paragraph{Theorem.}
(1) Let $\alpha$ be a Poisson bi-vector field in a domain of $R^d$.
The formula:

\be
f\star{g}:=\sum\limits_{n=0}\limits^{\infty}{\hbar^n}\sum\limits_{\Gamma
\in{G_n}}
\omega_\Gamma{B_{\Gamma,\alpha}}(f,g)
\ee
defines an associative product. (2)
Changing coordinates, one obtains a gauge equivalent star product.
\vskip 6mm
In formula (6) expressions $B_{\Gamma,\alpha}(f,g)$ are constructed with the
help of
Kontsevich's diagrams $\Gamma\in{G_n}$, $G_n$ being a set of the
$n^{th}$-order diagrams and $\omega_\Gamma$ are constant coefficients
corresponding
to diagrams $\Gamma\in{G_n}$. The values of these coefficients can be computed
by the prescription given in [1].
\vskip 6mm
Up to the second order, this formula can be written as follows

\be
f*g=fg+\hbar{\alpha^{ab}\partial_a{f}\partial_b{g}}+\
+\frac{1}{2}\hbar^2\alpha^{ab}\alpha^{cd}{\partial_a\partial_c{f}\partial_b
\partial_d{g}}+\\
+\frac{1}{3}\hbar^{2}{\alpha^{as}\partial_s\alpha^{bc}}
(\partial_a\partial_b{f}\partial_c{g}+\partial_a\partial_b{g}\partial_c{f})
+O(\hbar^{3})
\ee
\vskip 8mm
\paragraph{5. Purpose of the paper.}

We are going to make a direct check of Kontsevich's statement that
changing coordinates one obtains a gauge equivalent star
product, up to the $\hbar^3$-order starting from the constant bi-vector
(in this case, the star product is Moyal's one).
For the symplectic case, this fact was stated in [2].

We make a change of variables in the Moyal product, which is nothing but
Kontsevich's star product with the
constant Poisson bi-vector (let us denote it
$\vartheta^{ij}$), and try to represent it with a help of gauge
transformations in the form  of (6)
in new coordinates $z^a(x^i)$ using only the Poisson bi-vector field
$\alpha^{ab}(\hbar)$
${\alpha_0}^{ab}=\vartheta^{ij}
\frac{\partial{z^a}}{\partial{x^i}}\frac{\partial{z^b}}{\partial{x^j}}$.
The result should be compared with Kontsevich's formula for non-constant
(non-Moyal) $\vartheta^{ij}$.
In this way, we obtain formula (8). In the third order, we see that bi-vector
terms can not
be rewritten in new coordinates and thus are interpreted as $\alpha_2^{ab}$
in formula (4):
$\alpha^{ab}(\hbar)=\sum\limits_{i\geq{0}}{\hbar}^{i}\alpha_i^{ab}$.
The definition of $\alpha_2$ is not unique as we
may add to the formula (8) a bi-vector term (for example,
${\partial_s}\alpha^{pt}\partial_p
\alpha^{so}\partial_o\partial_t{\alpha^{ab}}\partial_a{f}\partial_b{g}$)
and thus subtract it from $\alpha_2$. However, this kind of terms always
correspond to loop diagrams. If one considers the case when there are no
loops, this consequently fixes the value of
$\alpha$. In this case, the answer we are going to get is given by the
following formula

\be
f*g=fg+\hbar{\alpha^{ab}\partial_a{f}\partial_b{g}}+\\
+\hbar^2\left[\frac{1}{2}
\alpha^{ab}\alpha^{cd}{\partial_a\partial_c{f}\partial_b\partial_d{g}}+
\frac{1}{3}{\alpha^{as}\partial_s\alpha^{bc}}
(\partial_a\partial_b{f}\partial_c{g}+\partial_a\partial_b{g}\partial_c{f})
\right]+
\\
+\hbar^{3}\left[\frac{1}{6}
\alpha^{ab}\alpha^{cd}\alpha^{ho}\partial_a\partial_c\partial_h{f}
\partial_b\partial_d\partial_o{g}+
\frac{1}{3}{\alpha^{tp}\partial_p\alpha^{as}\partial_s\partial_t\alpha^{bc}}
(\partial_a\partial_c{f}\partial_b{g}-\partial_a\partial_c{g}\partial_b{f})+
\right.
\\
+[\frac{2}{3}\alpha^{dp}\partial_{p}\alpha^{as}\partial_{s}\alpha^{bc}+
\frac{1}{3}\alpha^{ap}\partial_{p}\alpha^{ds}\partial_{s}\alpha^{cb}]
\partial_a\partial_c{f}\partial_b\partial_d{g}+\\
+\frac{1}{6}\alpha^{as}\alpha^{ct}\partial_s\partial_t\alpha^{bd}
(\partial_a\partial_b\partial_c{f}\partial_d{g}-
\partial_a\partial_b\partial_c{g}\partial_d{f})+
\\
\left.+\frac{1}{3}\alpha^{as}\partial_s\alpha^{bc}\alpha^{hd}
(\partial_a\partial_b\partial_h{f}\partial_c\partial_d{g}-
\partial_a\partial_b\partial_h{g}\partial_c\partial_d{f})\right]+O(\hbar^{4})
\ee
where

\be
\alpha^{ab}=\vartheta^{ij}
\frac{\partial{z^a}}{\partial{x^i}}\frac{\partial{z^b}}{\partial{x^j}}+
\hbar^{2}\left[\frac{1}{3!}\vartheta^{ij}\vartheta^{kl}\vartheta^{mn}
\frac{\partial^{3}{z^a}}{\partial{x^i}\partial{x^k}\partial{x^m}}
\frac{\partial^{3}{z^b}}{\partial{x^j}\partial{x^l}\partial{x^n}}-\right.\\
\left.-\frac{1}{18}S^{spt}\partial_p\partial_s\partial_t\alpha^{ab}
-\frac{1}{4}\vartheta^{ij}\vartheta^{kl}
\frac{\partial^2{z^s}}{\partial{x^i}\partial{x^k}}
\frac{\partial^2{z^t}}{\partial{x^j}\partial{x^l}}
\partial_s\partial_t{\alpha^{ab}}\right]+O(\hbar^3)
\ee
and $S^{abs}$ is determined in (15). The differential operator in the gauge
transformation (2) necessary for obtaining (8),(9) is the following

$$
D=1+\hbar^2\left[-\frac{1}{4}\vartheta^{ij}\vartheta^{kl}
\frac{\partial^{2}{z^a}}{\partial{x^i}\partial{x^k}}
\frac{\partial^{2}{z^b}}{\partial{x^j}\partial{x^l}}\partial_a\partial_b
+\right.
$$
$$
+\left.-\frac{1}{18}\vartheta^{ij}\vartheta^{kl}\left(
\frac{\partial^2{z^a}}{\partial{x^i}\partial{x^k}}
\frac{\partial{z^b}}{\partial{x^j}}\frac{\partial{z^c}}{\partial{x^l}}+
\frac{\partial^2{z^c}}{\partial{x^i}\partial{x^k}}
\frac{\partial{z^a}}{\partial{x^j}}\frac{\partial{z^b}}{\partial{x^l}}+
\frac{\partial^2{z^b}}{\partial{x^i}\partial{x^k}}
\frac{\partial{z^c}}{\partial{x^j}}\frac{\partial{z^a}}{\partial{x^l}}\right)
\partial_a\partial_b\partial_c \right]+O(\hbar^4)
$$
We show that $\alpha(\hbar)$ defined by (9) do not satisfy the Jacobi
identity: $\alpha^{as}\partial_s\alpha^{bc}+
\alpha^{cs}\partial_s\alpha^{ab}+\alpha^{bs}\partial_s\alpha{ca}=0$.
Still $\alpha_0$ does satisfy the Jacobi identity and we will use this fact
to compute coefficients in (8) from the requirement of associativity.
These coefficients are in agreement with [3]. It is discussed in section 5.
\vskip 8mm

\section{Diagram representation}

There is a natural way to represent separate terms in Kontsevich's formula by
diagrams.

For the $n^{th}$-order term one needs $n$ vertices, each vertex containing
$\alpha$,
and two more vertices containing functions $f$ and $g$. Vertices can be
connected by arrows. Each vertex is an origin of two ordered arrows.
If an arrow ends in some vertex, it describes a partial derivative acting on
this vertex.

The following diagrams correspond to formula (8)
(here we do not draw the arrows as all
of them point to the right):

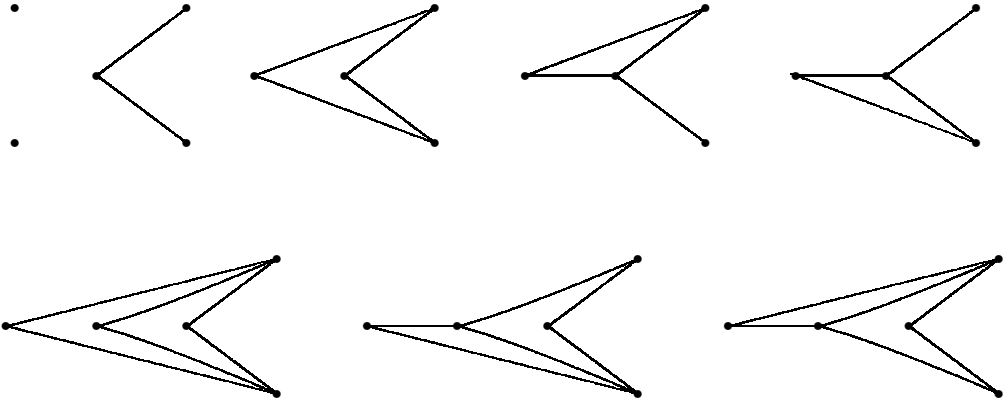
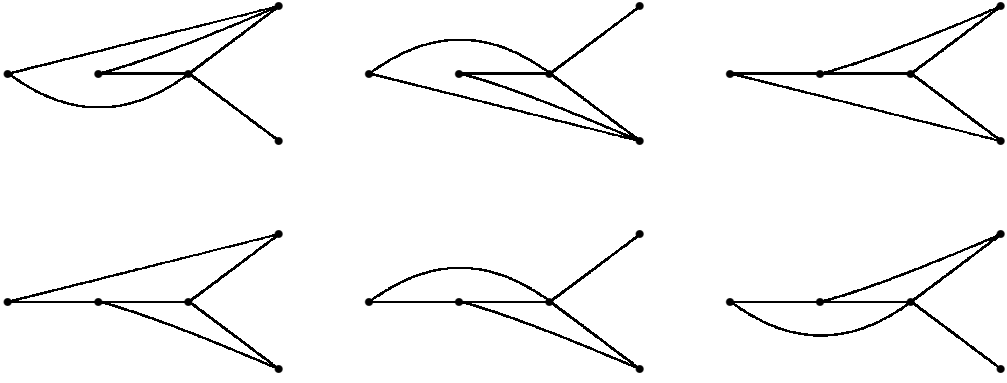
\setlength{\unitlength}{0.6mm}
\linethickness{0.4pt}
\begin{picture}(240,40)(0,0)
\put(2,5){\circle*{1.5}}
\put(2,35){\circle*{1.5}}

\put(20,20){\circle*{1.5}}
\put(40,5){\circle*{1.5}}
\put(40,35){\circle*{1.5}}
\bezier{200}(20,20)(24,23)(40,35)
\bezier{200}(20,20)(24,17)(40,5)

\put(55,20){\circle*{1.5}}
\put(75,20){\circle*{1.5}}
\put(95,5){\circle*{1.5}}
\put(95,35){\circle*{1.5}}
\bezier{200}(55,20)(63,17)(95,5)
\bezier{200}(55,20)(63,23)(95,35)
\bezier{200}(75,20)(79,23)(95,35)
\bezier{200}(75,20)(79,17)(95,5)

\put(115,20){\circle*{1.5}}
\put(135,20){\circle*{1.5}}
\put(155,5){\circle*{1.5}}
\put(155,35){\circle*{1.5}}
\bezier{200}(115,20)(125,20)(135,20)
\bezier{200}(115,20)(123,23)(155,35)
\bezier{200}(135,20)(139,23)(155,35)
\bezier{200}(135,20)(139,17)(155,5)

\put(175,20){\circle*{1.5}}
\put(195,20){\circle*{1.5}}
\put(215,5){\circle*{1.5}}
\put(215,35){\circle*{1.5}}
\bezier{200}(175,20)(185,20)(195,20)
\bezier{200}(175,20)(167,23)(215,5)
\bezier{200}(195,20)(199,23)(215,35)
\bezier{200}(195,20)(199,17)(215,5)
\end{picture}

\vskip 6mm
\setlength{\unitlength}{.6mm}
\begin{picture}(255,40)(0,0)
\put(0,15){\circle*{1.5}}
\put(20,15){\circle*{1.5}}
\put(40,15){\circle*{1.5}}
\put(60,0){\circle*{1.5}}
\put(60,30){\circle*{1.5}}
\bezier{200}(0,15)(12,12)(60,0)
\bezier{200}(0,15)(12,18)(60,30)
\bezier{200}(20,15)(32,18)(60,30)
\bezier{200}(20,15)(32,12)(60,0)
\bezier{200}(40,15)(44,18)(60,30)
\bezier{200}(40,15)(44,12)(60,0)

\put(80,15){\circle*{1.5}}
\put(100,15){\circle*{1.5}}
\put(120,15){\circle*{1.5}}
\put(140,0){\circle*{1.5}}
\put(140,30){\circle*{1.5}}
\bezier{200}(80,15)(92,12)(140,0)
\bezier{200}(80,15)(90,15)(100,15)
\bezier{200}(100,15)(112,18)(140,30)
\bezier{200}(100,15)(112,12)(140,0)
\bezier{200}(120,15)(124,18)(140,30)
\bezier{200}(120,15)(124,12)(140,0)

\put(160,15){\circle*{1.5}}
\put(180,15){\circle*{1.5}}
\put(200,15){\circle*{1.5}}
\put(220,0){\circle*{1.5}}
\put(220,30){\circle*{1.5}}
\bezier{200}(160,15)(170,15)(180,15)
\bezier{200}(160,15)(172,18)(220,30)
\bezier{200}(180,15)(192,18)(220,30)
\bezier{200}(180,15)(192,12)(220,0)
\bezier{200}(200,15)(204,18)(220,30)
\bezier{200}(200,15)(204,12)(220,0)

\end{picture}

\vskip 6mm

\begin{picture}(250,40)(0,0)
\put(0,15){\circle*{1.5}}
\put(20,15){\circle*{1.5}}
\put(40,15){\circle*{1.5}}
\put(60,0){\circle*{1.5}}
\put(60,30){\circle*{1.5}}
\bezier{200}(0,15)(12,18)(60,30)
\bezier{200}(20,15)(30,15)(40,15)
\bezier{200}(20,15)(32,18)(60,30)
\bezier{200}(0,15)(20,0)(40,15)
\bezier{200}(40,15)(44,18)(60,30)
\bezier{200}(40,15)(44,12)(60,0)

\put(80,15){\circle*{1.5}}
\put(100,15){\circle*{1.5}}
\put(120,15){\circle*{1.5}}
\put(140,0){\circle*{1.5}}
\put(140,30){\circle*{1.5}}
\bezier{200}(100,15)(110,15)(120,15)
\bezier{200}(80,15)(92,12)(140,0)
\bezier{200}(100,15)(112,12)(140,0)
\bezier{200}(80,15)(100,30)(120,15)
\bezier{200}(120,15)(124,18)(140,30)
\bezier{200}(120,15)(124,12)(140,0)

\put(160,15){\circle*{1.5}}
\put(180,15){\circle*{1.5}}
\put(200,15){\circle*{1.5}}
\put(220,0){\circle*{1.5}}
\put(220,30){\circle*{1.5}}
\bezier{200}(160,15)(170,15)(180,15)
\bezier{200}(180,15)(190,15)(200,15)
\bezier{200}(160,15)(172,12)(220,0)
\bezier{200}(180,15)(192,18)(220,30)
\bezier{200}(200,15)(204,18)(220,30)
\bezier{200}(200,15)(204,12)(220,0)
\end{picture}

\vskip 6mm

\begin{picture}(255,40)(0,0)
\put(0,15){\circle*{1.5}}
\put(20,15){\circle*{1.5}}
\put(40,15){\circle*{1.5}}
\put(60,0){\circle*{1.5}}
\put(60,30){\circle*{1.5}}
\bezier{200}(0,15)(10,15)(20,15)
\bezier{200}(20,15)(30,15)(40,15)
\bezier{200}(0,15)(12,18)(60,30)
\bezier{200}(20,15)(32,12)(60,0)
\bezier{200}(40,15)(44,18)(60,30)
\bezier{200}(40,15)(44,12)(60,0)

\put(80,15){\circle*{1.5}}
\put(100,15){\circle*{1.5}}
\put(120,15){\circle*{1.5}}
\put(140,0){\circle*{1.5}}
\put(140,30){\circle*{1.5}}\
\bezier{200}(100,15)(110,15)(120,15)
\bezier{200}(80,15)(90,15)(100,15)
\bezier{200}(100,15)(112,12)(140,0)
\bezier{200}(80,15)(100,30)(120,15)
\bezier{200}(120,15)(124,18)(140,30)
\bezier{200}(120,15)(124,12)(140,0)

\put(160,15){\circle*{1.5}}
\put(180,15){\circle*{1.5}}
\put(200,15){\circle*{1.5}}
\put(220,0){\circle*{1.5}}
\put(220,30){\circle*{1.5}}
\bezier{200}(180,15)(190,15)(200,15)
\bezier{200}(160,15)(170,15)(180,15)
\bezier{200}(180,15)(192,18)(220,30)
\bezier{200}(160,15)(180,0)(200,15)
\bezier{200}(200,15)(204,18)(220,30)
\bezier{200}(200,15)(204,12)(220,0)
\end{picture}

\vskip 6mm
Note that there are no loop diagrams in this picture. In particular, there is no
structure of the type $\partial_c\alpha^{da}\partial_d\alpha^{bc}$.

\section{Calculations up to the second order terms}

Let us consider the terms up to the second order ($\alpha=\alpha_0$)
in the Moyal formula (5)
and make a change of variables in it.
We are going to obtain formula (7) with the help of gauge transformations.
The change of variables gives the following expression for the Moyal product

\be
f\star{g}=fg+\hbar\alpha^{ab}\partial_a{f}\partial_b{g}+
\\+\frac{1}{2}\hbar^2
\vartheta^{ij}\vartheta^{kl}\left(
\frac{\partial^{2}{z^a}}{\partial{x^i}\partial{x^k}}
\frac{\partial{f}}{\partial{z^a}}+\frac{\partial{z^a}}{\partial{x^i}}
\frac{\partial^2{f}}{\partial{z^a}\partial{z^c}}
\frac{\partial{z^c}}{\partial{x^i}}\right)
\left(\frac{\partial^2{z^b}}{\partial{x^j}
\partial{x^l}}
\frac{\partial{g}}{\partial{z^b}}+\frac{\partial{z^b}}{\partial{x^l}}
\frac{\partial^2{g}}{\partial{z^b}\partial{z^d}}
\frac{\partial{z^d}}{\partial{x^j}}\right)=\\
=fg+\hbar\alpha^{ab}\partial_a{f}\partial_b{g}+
\frac{1}{2}\hbar^2\alpha^{ab}\alpha^{cd}
\partial_a\partial_c{f}\partial_b\partial_d{g}+\\
+\frac{1}{2}\hbar^2\vartheta^{ij}\vartheta^{kl}
\frac{\partial^{2}{z^a}}{\partial{x^i}\partial{x^k}}
\frac{\partial^{2}{z^b}}{\partial{x^j}\partial{x^l}}\partial_a{f}\partial_b{g}+
\\
+\hbar^2\vartheta^{ij}\vartheta^{kl}\frac{\partial^2{z^c}}{\partial{x^i}
\partial{x^k}}
\frac{\partial{z^a}}{\partial{x^l}}\frac{\partial{z^b}}{\partial{x^j}}
(\partial_a\partial_b{f}\partial_c{g}+\partial_a\partial_b{g}\partial_c{f})
+O(\hbar^3)
\ee
The last term is a symmetric bi-vector and can be canceled by the following
gauge
transformation

\be
{D'}_2=-\frac{1}{4}\vartheta^{ij}\vartheta^{kl}
\frac{\partial^{2}{z^a}}{\partial{x^i}\partial{x^k}}
\frac{\partial^{2}{z^b}}{\partial{x^j}\partial{x^l}}\partial_a\partial_b
\ee
Now we pay attention only to the terms $\partial^2{f}\partial{g}$,
$\partial^2{g}\partial{f}$ which are

\be
\frac{1}{2}P_1=\frac{1}{2}\frac{\partial^2{z^c}}{\partial{x^i}
\partial{x^k}}\frac{\partial{z^a}}{\partial{x^l}}
\frac{\partial{z^b}}{\partial{x^j}}
(\partial_a\partial_b{f}\partial_c{g}+\partial_a\partial_b{g}\partial_c{f})
\ee
If one puts $\alpha^{ab}=
\vartheta^{ij}\frac{\partial{z^a}}{\partial{x^i}}
\frac{\partial{z^b}}{\partial{x^j}}$
in $K_0{\alpha^{as}\partial_s\alpha^{bc}}
(\partial_a\partial_b{f}\partial_c{g}+\partial_a\partial_b{g}\partial_c{f})
$,
where $K_0$ is constant, this gives

\be
K_0\left(-\frac{\partial^2{z^a}}{\partial{x^i}
\partial{x^k}}\frac{\partial{z^c}}{\partial{x^l}}
\frac{\partial{z^b}}{\partial{x^j}}
+\frac{\partial^2{z^c}}{\partial{x^i}
\partial{x^k}}\frac{\partial{z^a}}{\partial{x^l}}
\frac{\partial{z^b}}{\partial{x^j}}\right)
(\partial_a\partial_b{f}\partial_c{g}+\partial_a\partial_b{g}\partial_c{f})=
K_0{P_1}-K_0{P_2}
\ee
where $P_2=\frac{\partial^2{z^a}}{\partial{x^i}
\partial{x^k}}\frac{\partial{z^c}}{\partial{x^l}}
\frac{\partial{z^b}}{\partial{x^j}}
(\partial_a\partial_b{f}\partial_c{g}+\partial_a\partial_b{g}\partial_c{f})$.

We are going now to find a gauge transformation that makes (12)
equal to expression (13). Consider

\be
D_2=K_1S^{abc}\partial_a\partial_b\partial_c
\ee
\be
S^{abc}=\vartheta^{ij}\vartheta^{kl}
\left(\frac{\partial^2{z^a}}{\partial{x^i}\partial{x^k}}
\frac{\partial{z^b}}{\partial{x^j}}\frac{\partial{z^c}}{\partial{x^l}}+
\frac{\partial^2{z^c}}{\partial{x^i}\partial{x^k}}
\frac{\partial{z^a}}{\partial{x^j}}\frac{\partial{z^b}}{\partial{x^l}}+
\frac{\partial^2{z^b}}{\partial{x^i}\partial{x^k}}
\frac{\partial{z^c}}{\partial{x^j}}\frac{\partial{z^a}}{\partial{x^l}}\right)
\ee
Note that $S^{abc}$ is symmetric with respect to $a,b,c$.
Transformation (14-15) adds to (12) the following terms

\be
D_2(fg)-fD_2(g)-gD_2(f)=3S^{abc}
(\partial_a\partial_b{f}\partial_c{g}+\partial_a\partial_b{g}\partial_c{f})=
3K_1(P_1+2P_2)
\ee
The equality $(12)+(16)=(13)$ looks like

$$\frac{1}{2}P_1+3K_1(P_1+2P_2)=K_0(P_1-P_2)$$
$$K_0=\frac{1}{3},K_1=-\frac{1}{18}$$

This means we should substitute in formula (3)
the gauge transformations ${D'}_2$ (11) and $D_2$ (14-15),
where $K_1=-\frac{1}{18}$. Note that one can add to the star product
in formula (3) the symmetric bi-vector

\be
\lambda\hbar^2\partial_d{\alpha^{ac}}
\partial_c{\alpha^{bd}}
\ee
(where $\lambda$ is a constant)
which corresponds to loop diagram.

It is easy to get corrections to the expression for $B_3(f,g)$
from formula (1)
under gauge transformations $D_2$ made in the second order:

\be
B'_3(f,g)=B_3(f,g)+D_2B_1(f,g)-B_1(D_2f,g)-B_1(f,D_2g)
\ee
Thus, it is possible to change higher order terms by changing $\lambda$.
We will put $\lambda=0$.

\section{Calculations up to the third order terms}

Let us write down the terms of the third order from the Moyal product in the new
coordinates:

\be
\frac{1}{6}\vartheta^{ij}\vartheta^{kl}\vartheta^{mn}
\left(\frac{\partial^3{f}}{\partial{z^a}\partial{z^c}\partial{z^e}}
\frac{\partial{z^e}}{\partial{x^i}}\frac{\partial{z^c}}{\partial{x^k}}
\frac{\partial{z^a}}{\partial{x^m}}+
\frac{\partial^2{f}}{\partial{z^a}\partial{z^c}}
\left(\frac{\partial^2{z^a}}{\partial{x^k}\partial{x^m}}\frac{\partial{z^c}}
{\partial{x^i}}+
\frac{\partial^2{z^a}}{\partial{x^i}\partial{x^m}}
\frac{\partial{z^c}}{\partial{x^k}}+
\right.\right.\\
 \left.\left.+
\frac{\partial^2{z^a}}{\partial{x^i}\partial{x^k}}
\frac{\partial{z^c}}{\partial{x^m}}
\right)+
\frac{\partial^3{z^a}}{\partial{x^i}\partial{x^k}\partial{x^m}}
\frac{\partial{f}}{\partial{z^a}}\right)
\left(\frac{\partial^3{g}}{\partial{z^b}\partial{z^d}\partial{z^e}}
\frac{\partial{z^e}}{\partial{x^j}}\frac{\partial{z^d}}{\partial{x^l}}
\frac{\partial{z^b}}{\partial{x^n}}+\right.\\
\left.+\frac{\partial^2{g}}{\partial{z^b}\partial{z^d}}
\left(\frac{\partial^2{z^b}}{\partial{x^l}\partial{x^n}}
\frac{\partial{z^d}}{\partial{x^j}}+
\frac{\partial^2{z^b}}{\partial{x^l}\partial{x^j}}
\frac{\partial{z^d}}{\partial{x^n}}+
\frac{\partial^2{z^b}}{\partial{x^n}\partial{x^j}}
\frac{\partial{z^d}}{\partial{x^l}}\right)+
\frac{\partial^3{z^b}}{\partial{x^j}\partial{x^l}\partial{x^n}}
\frac{\partial{g}}{\partial{z^b}}\right)
\ee
Due to (18) we are to add to the above expression

\be
D_2B_1(f,g)-B_1(D_2f,g)-B_1(f,D_2g)+D'_2B_1(f,g)-B_1(D'_2f,g)-B_1(f,D'_2g)
\ee
where $D_2$ and $D'_2$ are given by (14-15) and (11).

Again we are going to represent the sum of the Moyal product and gauge terms
in diagrams (in the new coordinates).

An important remark is that there is no necessity to make gauge transformations
in
the third order, since the corrections
to the third order terms $B_3(f,g)$ coming from
the gauge transformation of the second order (20) are enough to represent
this expression in the form of Kontsevich's diagrams.

Now consider an example of computation for the terms of the type
$(\partial^2{f}\partial^2{g})$.
Such terms in the Moyal product (19) are

\be
\frac{1}{6}\vartheta^{ij}\vartheta^{kl}\vartheta^{mn}
\left(\frac{\partial^2{z^a}}{\partial{x^k}
\partial{x^m}}\frac{\partial{z^c}}{\partial{x^i}}+
\frac{\partial^2{z^a}}{\partial{x^i}\partial{x^m}}
\frac{\partial{z^c}}{\partial{x^k}}+
\frac{\partial^2{z^a}}{\partial{x^i}\partial{x^k}}
\frac{\partial{z^c}}{\partial{x^m}}\right)
\left(\frac{\partial^2{z^b}}{\partial{x^l}
\partial{x^n}}\frac{\partial{z^d}}{\partial{x^j}}+\right.\\
\left.+
\frac{\partial^2{z^b}}{\partial{x^l}\partial{x^j}}
\frac{\partial{z^d}}{\partial{x^n}}+
\frac{\partial^2{z^b}}{\partial{x^n}
\partial{x^j}}\frac{\partial{z^d}}{\partial{x^l}}\right)
\partial_a\partial_c{f}\partial_b\partial_d{g}=\\
=\vartheta^{ij}\vartheta^{kl}\vartheta^{mn}\left(\frac{1}{2}
\frac{\partial^2{z^a}}{\partial{x^i}\partial{x^k}}
\frac{\partial^2{z^b}}{\partial{x^j}\partial{x^l}}
\frac{\partial{z^c}}{\partial{x^m}}\frac{\partial{z^d}}{\partial{x^n}}+
\frac{\partial^2{z^a}}{\partial{x^i}\partial{x^k}}
\frac{\partial^2{z^b}}{\partial{x^l}\partial{x^n}}
\frac{\partial{z^c}}{\partial{x^m}}\frac{\partial{z^d}}{\partial{x^j}}\right)
\partial_a\partial_c{f}\partial_b\partial_d{g}
\ee
The gauge terms are

\be
D'_2:-\frac{1}{2}\vartheta^{ij}\vartheta^{kl}\vartheta^{mn}
\frac{\partial^2{z^a}}{\partial{x^i}\partial{x^k}}
\frac{\partial^2{z^b}}{\partial{x^j}\partial{x^l}}
\frac{\partial{z^c}}{\partial{x^m}}\frac{\partial{z^d}}{\partial{x^n}}
\partial_a\partial_c{f}\partial_b\partial_d{g}
\ee
(Note that this gauge transformation cancels the first term in the right side of
(21).)

\be
D_2: -\frac{1}{3}S^{scd}\partial_s\alpha^{ab}
\partial_a\partial_c{f}\partial_b\partial_d{g}
\ee
where $S^{scd}$ is defined by (15). One may check that (23) is equal
to

\be
 -\frac{1}{3}S^{scd}\partial_s\alpha^{ab}
\partial_a\partial_c{f}\partial_b\partial_d{g}=
-\frac{1}{3}Y_1-\frac{1}{3}Y_2-\frac{1}{3}Y_3
\ee
where $Y_1$,$Y_2$ and $Y_3$ are

\be
Y_1=\vartheta^{ij}\vartheta^{kl}\vartheta^{mn}
\frac{\partial{z^a}}{\partial{x^j}}
\frac{\partial^2{z^d}}{\partial{x^k}\partial{x^i}}
\left(\frac{\partial^2{z^c}}{\partial{x^m}\partial{x^l}}
\frac{\partial{z^b}}{\partial{x^n}}-
\frac{\partial^2{z^b}}{\partial{x^m}\partial{x^l}}
\frac{\partial{z^c}}{\partial{x^n}}
\right)
\ee
\be
Y_2=\vartheta^{ij}\vartheta^{kl}\vartheta^{mn}
\frac{\partial{z^a}}{\partial{x^j}}
\frac{\partial{z^d}}{\partial{x^l}}
\frac{\partial^2{z^s}}{\partial{x^i}\partial{x^k}}
\frac{\partial{x^o}}{\partial{z^s}}
\left(\frac{\partial^2{z^c}}{\partial{x^m}\partial{x^o}}
\frac{\partial{z^b}}{\partial{x^n}}-
\frac{\partial^2{z^b}}{\partial{x^m}\partial{x^o}}
\frac{\partial{z^c}}{\partial{x^n}}
\right)
\ee
\be
Y_3=\vartheta^{ij}\vartheta^{kl}\vartheta^{mn}
\frac{\partial{z^d}}{\partial{x^j}}
\frac{\partial^2{z^a}}{\partial{x^k}\partial{x^i}}
\left(\frac{\partial^2{z^c}}{\partial{x^m}\partial{x^l}}
\frac{\partial{z^b}}{\partial{x^n}}-
\frac{\partial^2{z^b}}{\partial{x^m}\partial{x^l}}
\frac{\partial{z^c}}{\partial{x^n}}
\right)
\ee
We will try to find an answer in the form

\be
[A\alpha^{dp}\partial_{p}\alpha^{as}\partial_{s}\alpha^{bc}+
B\alpha^{ap}\partial_{p}\alpha^{ds}\partial_{s}\alpha^{cb}]
\partial_a\partial_c{f}\partial_b\partial_d{g}
\ee
In old coordinates it looks like

\be
[A\alpha^{dp}\partial_{p}\alpha^{as}\partial_{s}\alpha^{bc}+
B\alpha^{ap}\partial_{p}\alpha^{ds}\partial_{s}\alpha^{cb}]
\partial_a\partial_c{f}\partial_b\partial_d{g}=A(Y_2-Y_1)+B(Y_3-Y_2)
\ee
where $A$ and $B$ are constants. Note also that the last term in the
right part of (21) is equal to $Y_3$ since

$$\vartheta^{ij}\vartheta^{kl}\vartheta^{mn}
\frac{\partial^2{z^a}}{\partial{x^i}\partial{x^k}}
\frac{\partial^2{z^c}}{\partial{x^l}\partial{x^n}}
\frac{\partial{z^b}}{\partial{x^m}}\frac{\partial{z^d}}{\partial{x^j}}
\partial_a\partial_c{f}\partial_b\partial_d{g}=0$$
due to the skew symmetry with respect to the interchange $(a\leftrightarrow{c}$
and
$d\leftrightarrow{b})$.
Now we can find $A$ and $B$

$$Y_3-\frac{1}{3}Y_1-\frac{1}{3}Y_2-\frac{1}{3}Y_3=A(Y_3-Y_2)+B(Y_2-Y_1)$$
$$A=\frac{2}{3},\ B=\frac{1}{3}$$
Thus, we have found the $(\partial^2{f}\partial^2{g})$-terms in the deformation
quantization formula (8):

$$\hbar^{3}[\frac{2}{3}\alpha^{dp}\partial_{p}\alpha^{as}\partial_{s}
\alpha^{bc}+
\frac{1}{3}\alpha^{ap}\partial_{p}\alpha^{ds}\partial_{s}\alpha^{cb}]
\partial_a\partial_c{f}\partial_b\partial_d{g}$$
All other terms can be found in the same manner.  The only exception
happens with the $\partial{f}\partial{g}$-terms. For these terms, one has

\be
\hbox{Moyal term}: \frac{1}{6}\vartheta^{ij}\vartheta^{kl}\vartheta^{mn}
\frac{\partial^3{z^a}}{\partial{x^i}\partial{x^k}\partial{x^m}}
\frac{\partial^3{z^b}}{\partial{x^j}\partial{x^l}\partial{x^n}}
\ee
\be
\hbox{Gauge}\ D_2: -\frac{1}{18}S^{spt}\partial_p\partial_s\partial_t\alpha^{ab}
\partial_a{f}\partial_b{g}
\ee
\be
\hbox{Gauge}\ D'_2: -\frac{1}{4}\vartheta^{ij}\vartheta^{kl}
\frac{\partial^2{z^s}}{\partial{x^i}\partial{x^k}}
\frac{\partial^2{z^t}}{\partial{x^j}\partial{x^l}}
\partial_s\partial_t{\alpha^{ab}}\partial_a{f}\partial_b{g}
\ee
These terms can not be canceled by gauge transformations as they are
skew symmetric, and none of them can be represented in the form of
Kontsevich's
diagrams. Therefore, we are going to treat them as ${\alpha_2}^{ab}\partial_a{f}
\partial_b{g}$.
The definition of $\alpha_2$ is not unique for the following reason.
Formula (32) is contained in
$\partial_p\alpha^{so}\partial_o\alpha^{tp}\partial_s\partial_t
\alpha^{ab}\partial_a{f}\partial_b{g}$, however, this expression contains
other terms. Formula (31) can not be realized in the form of
Kontsevich's
diagrams at all because of the third derivative. The Moyal term (30) contains
$\partial^3{z^a}\partial^3{z^b}$ and thus is contained in the only
diagram: $\alpha^{pt}\partial_t\partial_s\alpha^{ao}
\partial_o\partial_p\alpha^{bs}$ but it again contains many other terms.

Thus we have following expression for the Poisson bi-vector

\be
\alpha^{ab}=\vartheta^{ij}
\frac{\partial{z^a}}{\partial{x^i}}\frac{\partial{z^b}}{\partial{x^j}}+
\hbar^{2}\left[\frac{1}{3!}\vartheta^{ij}\vartheta^{kl}\vartheta^{mn}
\frac{\partial^{3}{z^a}}{\partial{x^i}\partial{x^k}\partial{x^m}}
\frac{\partial^{3}{z^b}}{\partial{x^j}\partial{x^l}\partial{x^n}}-\right.\\
\left.-\frac{1}{18}S^{spt}\partial_p\partial_s\partial_t\alpha^{ab}
-\frac{1}{4}\vartheta^{ij}\vartheta^{kl}
\frac{\partial^2{z^s}}{\partial{x^i}\partial{x^k}}
\frac{\partial^2{z^t}}{\partial{x^j}\partial{x^l}}
\partial_s\partial_t{\alpha^{ab}}\right]+O(\hbar^3)
\ee
So we found $\alpha=\alpha_0+\hbar^2\alpha_2$. It is obvious that
$\alpha^{ab}=\vartheta^{ij}\frac{\partial{z^a}}{\partial{x^i}}
\frac{\partial{z^b}}{\partial{x^j}}$ satisfy the Jacobi identity

$${\alpha_0}^{as}\partial_s{{\alpha_0}^{bc}}+
{\alpha_0}^{cs}\partial_s{{\alpha_0}^{ab}}+
{\alpha_0}^{bs}\partial_s{{\alpha_0}^{ca}}=0 $$
The Poisson bi-vector $\alpha(\hbar)$ does not satisfy the Jacobi identity in
the $\hbar^2$-order.
This statement may be proved in the following way.
One may consider a certain type of terms in the right hand side of the Jacobi
identity and show that there is no way to drop them off.

In the $\hbar^2$-order, the Jacobi identity looks like

$${\alpha_0}^{as}\partial_s{{\alpha_2}^{bc}}+
{\alpha_2}^{as}\partial_s{{\alpha_0}^{bc}}
+{\alpha_0}^{cs}\partial_s{{\alpha_2}^{ab}}+
{\alpha_2}^{cs}\partial_s{{\alpha_0}^{ab}}
+{\alpha_0}^{bs}\partial_s{{\alpha_2}^{ca}}+
{\alpha_2}^{bs}\partial_s{{\alpha_0}^{ca}}=0$$
To see it is not true for $\alpha$ from (33) one should have a look on the terms
$\partial{z}\partial^4{z}\partial^3{z} $. The coordinate $z$ can have different
types of indices: external $a,b,c$ and internal $o,t,p$. Thus, there is a
number of the above expressions with different types of indices and each of them
should satisfy the Jacobi identity in order to have it for the whole
$\alpha$.
This kind of terms
appears from the two first terms in (33). The indices have the following
structures:
$\partial{z^a}\partial^4{z^b}\partial^3{z^c}$ from the first term,
$\partial{z^a}\partial^4{z^s}\partial^3{z^b}$ and
$\partial{z^a}\partial^4{z^b}\partial^3{z^s}$ from the second one.
Each of the above expressions makes the Jacobi identity non-valid.
Still, we should remember that $\alpha_2$ can be defined in a different way if
one
adds $\partial{f}\partial{g}$ terms to the star product (8). The only such a
term
which gives the same type of terms in the Jacobi identity is
$\partial_p{\alpha^{so}}\partial_o{\alpha^{tp}}\partial_s
\partial_t{\alpha^{ab}}$.
It can cancel $\partial{z^a}\partial^4{z^b}\partial^3{z^c}$ and add
$\partial{z^a}\partial^4{z^s}\partial^3{z^b}$,
$\partial{z^a}\partial^4{z^b}\partial^3{z^s}$ and
$\partial{z^a}\partial^4{z^o}\partial^3{z^s}$.
In this case, the expression $\partial{z^a}\partial^4{z^o}\partial^3{z^s}$
violates the Jacobi identity.
Note that we did not consider the "tadpole" diagrams (which contain
$\partial_s{\alpha^{sa}}$), since there are no such terms in Kontsevich's
formula but they save the Jacobi identity neither.

\section{Values of coefficients from associativity}
In this section, we are going to obtain the coefficients in formula (8)
from the condition of associativity and the Jacobi identity for $\alpha_0$.
Associativity for the $n^{th}$-order means $(f\star{g})\star{h}=
f\star(g\star{h})+
O(\hbar^{n+1})$. For the second and the third orders, we have
(star product is defined by (1)):

\be
B_2(fg,h)+B_1(B_1(f,g),h)+B_2(f,g)h=B_2(f,gh)+B_1(f,B_1(g,h))+fB_2(g,h)
\ee
\be
B_3(fg,h)+B_2(B_1(f,g),h)+B_1(B_2(f,g),h)+B_3(f,g)h=\\
   =B_3(f,gh)+B_2(f,B_1(g,h))+B_1(f,B_2(g,h))+fB_3(g,h)
\ee
In the second order condition (34), one may put $\alpha=\alpha_0$. In the
third order (35), $B_3(f,g)$ contains the term depending on $\alpha_2$, but
this is
a bi-vector ${\alpha_2}^{ab}\partial_a{f}\partial_b{g}$ and, thus,
cancels from (35) due to the Leibnitz rule. It means there is no possibility
to obtain some knowledge about $\alpha_2$ from associativity in the
$\hbar^3$-order.
So we may use the Jacobi identity for $\alpha$ in equations
(34),(35).

Let us put arbitrary coefficients in formula (8)

\be
f*g=fg+\hbar{\alpha^{ab}\partial_a{f}\partial_b{g}}+\\
+\hbar^2[A_1\alpha^{ab}\alpha^{cd}{\partial_a\partial_c{f}\partial_b
\partial_d{g}}+
A_2{\alpha^{as}\partial_s\alpha^{bc}}
\partial_a\partial_b{f}\partial_c{g}+
A_3{\alpha^{as}\partial_s\alpha^{bc}}\partial_a\partial_b{g}\partial_c{f}]+
\\
\hbar^3[C_1\alpha^{ab}\alpha^{cd}\alpha^{ho}\partial_a\partial_c\partial_h{f}
\partial_b\partial_d\partial_o{g}+
C_2{\alpha^{tp}\partial_p\alpha^{as}\partial_s\partial_t\alpha^{bc}}
\partial_a\partial_c{f}\partial_b{g}+
\\+C_3{\alpha^{tp}\partial_p\alpha^{as}\partial_s\partial_t\alpha^{bc}}
\partial_a\partial_c{g}\partial_b{f}+\\
+[C_4\alpha^{dp}\partial_{p}\alpha^{as}\partial_{s}\alpha^{bc}+
C_5\alpha^{ap}\partial_{p}\alpha^{ds}\partial_{s}\alpha^{cb}]
\partial_a\partial_c{f}\partial_b\partial_d{g}+
\\+C_6\alpha^{as}\alpha^{ct}\partial_s\partial_t\alpha^{bd}
\partial_a\partial_b\partial_c{f}\partial_d{g}+
C_7\alpha^{as}\alpha^{ct}\partial_s\partial_t\alpha^{bd}
\partial_a\partial_b\partial_c{g}\partial_d{f}+
\\
+C_8\alpha^{as}\partial_s\alpha^{bc}\alpha^{hd}
\partial_a\partial_b\partial_h{f}\partial_c\partial_d{g}+
C_9\alpha^{as}\partial_s\alpha^{bc}\alpha^{hd}
\partial_a\partial_b\partial_h{g}\partial_c\partial_d{f}]+O(\hbar^{4})
\ee
Equation (34) simply gives $A_1=\frac{1}{2}$ and

$$((1-A_2)\alpha^{as}\partial_s\alpha^{bc}+(1-A_3)\alpha^{cs}
\partial_s\alpha^{ab}+
(A_2+A_3)\alpha^{bs}\partial_s\alpha^{ca})\partial_a{f}\partial_b{g}
\partial_c{h}=0
$$
As the Jacobi identity is known to be satisfied by $\alpha_0$, one obtains

$$1-A_2=1-A_3=A_2+A_3$$
and thus $A_2=A_3=\frac{1}{3}$.

Similarly from equation (35) one may obtain the following results

$$C_1=\frac{1}{6},C_2=-C_3=\frac{1}{3},C_4-C_5=\frac{1}{3},C_6=-C_7=\frac{1}{6}
C_8=-C_9=\frac{1}{3}$$
These results are in agreement with formula (8). At the same time, they are
in agreement with [3]. The only difference is that the authors of [3] write
$C_4=-C_5=\frac{1}{6}$
and, in formula (8), $C_4=\frac{2}{3},\ C_5=\frac{1}{3}$. The both cases
satisfy the obtained condition $C_4-C_5=\frac{1}{3}$. However, in our case the
two coefficients $C_4$ and $C_5$
are not fixed since the two diagrams corresponding to the terms
$\partial^2{f}\partial^2{g}$
are dependent through the Jacobi identity for $\alpha_0$. Since the entire
$\alpha$
does not satisfy the Jacobi identity, these coefficients in formula (8) can not
be changed. We expect that our coefficients will better suit next
orders calculations.

\section{Conclusion}
In the present paper, we checked, up to the third order in $\hbar$ the
statement made
in [1] that changing coordinates in Kontsevich's
star product leads to gauge equivalent star products. The manifest calculations
were presented
starting from the star product with the constant Poisson bi-vector (giving the
Moyal product). In this
way, we obtained formula (8) for the deformation quantization
which is agreement with [3].
We also obtained the same result from the requirement of associativity
of the star product and the Jacobi identity for $\alpha_0$ (which is not valid
for
$\alpha(\hbar)$).

\section{Acknowledgments}
Author is grateful to A.Gerasimov, A.Losev, A.Mironov, K.Saraikin and
especially to A.Morozov for introducing the problem and discussions.
This work was partially supported by RFBR grant N00-02-16530 and
the program for support of the scientific schools 00-15-96557.

\end{document}